\documentclass[prb,11pt]{revtex4-1}

\usepackage{amsmath}    
\usepackage{graphicx}   
\usepackage{verbatim}   
\usepackage{subfigure}  

\begin{document}

\title{On the Raman Instability in Degenerate Relativistic Plasmas}
\author{G.T.Chanturia$^1$, V.I.Berezhiani$^{1,2}$, S.M.Mahajan$^3$}
\affiliation{$^1$ School of Physics, Free university of Tbilisi, Tbilisi 0159, Georgia}
\affiliation{$^2$ Andronikashvili Institute of Physics (TSU), Tbilisi 0177, Georgia}
\affiliation{$^3$ Institute for Fusion Studies, The University of Texas at Austin, Austin, Texas 78712, USA}


\begin{abstract}
The stimulated Raman scattering instability in a fully degenerate electron
plasma is studied applying relativistic hydrodynamic and Maxwell equations. We
demonstrated that the instability develops for weakly as well as strongly
relativistic degenerate plasma. It is shown that in the field of strong
radiation relativistically degenerate plasma effectively responses as in the
case of weak degeneracy.
\end{abstract}

\maketitle

Astrophysical objects contain very intense radiation sources with their
spectral content spanning from radio to $\gamma$-ray emission [\onlinecite{Begelman}].
Most types of stars, interstellar gas, active galactic nuclei, distant quasars
\textit{etc.} are either comprised by significant fraction of plasma, or are
immersed in a plasma environment [\onlinecite{Shapiro}]. The nature of Radiation
emitted by such plasmas is usually interpreted in terms of bremsstrahlung and
synchrotron mechanism. However, under certain conditions the collective
parametric effects as, for instance, stimulated Raman scattering (SRS)
instability could leave definite signature on the radiation spectra
[\onlinecite{Zhel}]. Within the framework of classical plasmas, the SRS has been
studied in a variety of systems like the laser-driven inertial confinement
fusion, particle acceleration and plasma heating, and for probing of the
plasma parameters in laboratory conditions [\onlinecite{Kruer}]-[\onlinecite{kagan}]. Linear
and nonlinear stages of SRS instability have been investigated for cold as
well as relativistic hot plasma for EM radiation of arbitrary intensity.
Possible role of SRS instability in astrophysical plasma is discussed in
[\onlinecite{Kaplan}],[\onlinecite{Yuri}] and references therein; it is shown that induced
scattering could significantly affect radiation from sources with high
brightness temperatures.

\ For the extremely high density plasmas pertinent to astrophysical objects
such as in the interior of white dwarfs, the neutron or the pre-supernova
stars and presumably, at the source of gamma ray bursts [\onlinecite{Shapiro}], the
classical approximation breaks down; the relevant plasma is degenerate
requiring Fermi-Dirac statistics for a correct description. The Fermi energy
of the degenerate electron gas is greater than the binding energy with atomic
nucleus and as a consequence all atoms are in an ionized state. The plasma
number density is believed to be in the range from $N_{0}=10^{26}cm^{-3}$ to
$N_{0}=10^{34}$ $cm^{-3}$ and the matter behaves as a weakly coupled
degenerate plasma provided that averaged interparticle distance is smaller
than the thermal de Broglie wavelength [\onlinecite{Landau}]. \ At such densities, the
degenerate electron gas must be treated relativistically even when its
"temperature" is nonrelativistic, in fact, even zero.

The Fermi energy associated with a typical particles in a degenerate electron
gas, $\epsilon_{F}=m_{e}c^{2}\left(  \gamma_{F}-1\right)  $ (the Fermi
$\gamma_{F}=\sqrt{1+p_{F}^{2}/m_{e}^{2}c^{2}},$ $p_{F}=m_{e}c\left(
n_{R}/n_{c}\right)  ^{1/3}$ [\onlinecite{Haas}]), exceeds the rest mass energy for
densities $n_{R}$ greater than the critical density $n_{c}=m_{e}^{3}c^{3}%
/3\pi^{2}\hbar^{3}=5.9\times10^{29}cm^{-3}$.

In studies of nonlinear self-interactions of high frequency EM and plasma
waves in relativistic degenerate electron (as well as electron-positron)
plasmas [\onlinecite{Guga}]-[\onlinecite{Nana}], it has been shown such plasmas can support
stable localized EM structures for arbitrary level of degeneracy. The dynamics
of these nonlinear EM structures could provides a theoretical basis for
establishing the nature of the observed radiation. To the best of our
knowledge, the SRS instability of EM radiation in degenerate relativistic
plasma is not addressed so far. The SRS instability is a nonlinear parametric
process when a powerful electromagnetic (EM) wave decays into a plasma wave
and an EM wave. In an underdense plasma with $\omega_{0}>2\omega_{e0}$, where
$\omega_{0}$ is the mean frequency of carrier EM pulses and $\omega
_{e0}=\left(  4\pi e^{2}N_{0}/m_{e}\right)  ^{1/2}$ \ is the electron plasma
frequency, the SRS results in two EM sidebands upshifted and downshifted by
the plasma frequency. For a relativistic degenerate electron gas, the
radiation frequency at which SRS could be relevant, falls into soft or even
hard $X$-ray band. Consequently, it could leave definite footprints in the
spectra of astrophysical $X$-ray sources.

Our study of the SRS instability is based on the Maxwell equations and a
relativistic electron plasma fluid model. We will work In terms of the
familiar vectorial form of the fully covariant equations (see, for example
[\onlinecite{SM03}]-[\onlinecite{SMFA17}]) that translate as (in the Coulomb gauge
$\mathbf{\nabla\cdot A}=0$)%

\begin{equation}
\frac{\partial^{2}\mathbf{A}}{\partial t^{2}}-c^{2}\Delta\mathbf{A+}%
c\frac{\partial}{\partial t}\left(  \mathbf{\nabla}\varphi\right)  -4\pi
c\mathbf{J}=0\label{G1}%
\end{equation}%
\begin{equation}
\Delta\varphi=4\pi e(n_{R}\gamma-N_{0})\label{G2}%
\end{equation}
where the vector potential $\mathbf{A}$ and the scalar potential $\varphi$ are
the appropriate components of the EM four vector $A^{\mu}=[\varphi
,\mathbf{A}]$, and $\mathbf{J=-}en_{R}\mathbf{p}/m_{e}$, and $-ec\gamma n_{R}$
constitute the electron four current $J^{\mu}$.

The ions just provide a neutralizing background, and have a density $N_{0}$ in
their rest frame which is also taken to be the fiduciary/laboratory frame. The
"rest" frame electron density, $n_{R}$, is a Lorentz scalar and is related to
the laboratory frame density $N$ by the relation $N=\gamma n_{R}$. The
relativistic $\gamma=\sqrt{1+\mathbf{p}^{2}/m_{e}^{2}c^{2}}$ is determined
from the vector part $\mathbf{p}$ of the four momentum.

In the plasma with zero generalized vorticity $\mathbf{\Omega=\nabla\times
}\left(  G\mathbf{p-}e\mathbf{A/}c\right)  =0$ the fluid equations for the
electrons can be written as (see [\onlinecite{Guga}] for details):%

\begin{equation}
\frac{\partial}{\partial t}\left(  G\mathbf{p-}\frac{e}{c}\mathbf{A}\right)
+\mathbf{\nabla}\left(  m_{e}c^{2}G\gamma-e\varphi\right)  =0 \label{G3}%
\end{equation}

\begin{equation}
\frac{\partial}{\partial t}n_{R}\gamma+\mathbf{\nabla\cdot}\left(
n_{R}\mathbf{p}/m_{e}\right)  =0 \label{G4}%
\end{equation}
In Eq.(\ref{G3}), the "thermal" mass enhancement factor $G$ is defined as
$G=w/\left(  n_{R}m_{e}c^{2}\right)  $, where $w$ is the enthalpy per unit
volume. The effective mass factor is, generally, a non-trivial function of
\ plasma temperature and density for both classical Maxwell--Juttner, or the
quantum Fermi-Juttner statistics. However, if the thermal energy of the plasma
electrons is much lower than the Fermi energy, the plasma may be treated as
cold, (i.e. having zero temperature). The zero temperature approximation will
be adequate even for temperatures of order $10^{9}K$ allowing a particularly
simple expression, $G=\gamma_{F}=\sqrt{1+\left(  n_{R}/n_{c}\right)  ^{2/3}}$
[\onlinecite{Shapiro}].

For an EM wave propagating in the $z$ direction(all dynamic variables vary
only in $z$ and time $t$), the vector potential has just transverse components
$\mathbf{A}=(\mathbf{A}_{\perp},0)$. For this simplified 1-D propagation,
Eq.(\ref{G3}) can be readily integrated to yield%

\begin{equation}
\mathbf{p}_{\mathbf{\perp}}\mathbf{=}\frac{\mathbf{A}_{\perp}}{\gamma_{_{F}}}
\label{G5}%
\end{equation}
where%

\begin{equation}
\gamma_{F}\mathbf{=}\sqrt{1+\left(  R_{0}n\right)  ^{2/3}} \label{G6}%
\end{equation}
and \ %

\begin{equation}
\gamma\mathbf{=}\sqrt{1+\frac{\mathbf{A}_{\perp}^{2}}{\gamma_{F}^{2}}%
+p_{z}^{2}} \label{G7}%
\end{equation}
in terms of the normalized variables $\mathbf{A}_{\perp}=\left(
e\mathbf{A}_{\perp}/m_{e}c^{2}\right)  $, $\varphi=\left(  e\varphi/m_{e}%
c^{2}\right)  $, $\mathbf{p}=\left(  \mathbf{p}/mc\right)  $ and $n=\left(
n_{R}/N_{0}\right)  $.

The parameter $R_{0}=\left(  N_{0}/n_{c}\right)  $ measures the strength of
plasma degeneracy: for $R_{0}<<1$ the plasma is nonrelativistic, while for
$R_{0}\geq1$ plasma is in the relativistic degenerate state. Note that the
dimensionless Fermi momentum now reads as $p_{F}=\left(  R_{0}n\right)
^{1/3}$.

The Maxwell \ Eq.(\ref{G1}) is reduced to the wave equation
\begin{equation}
\frac{\partial^{2}\mathbf{A}_{\perp}}{\partial t^{2}}-c^{2}\frac{\partial
^{2}\mathbf{A}_{\perp}}{\partial z^{2}}\mathbf{+}\Omega_{e}^{2}\mathbf{A}%
_{\perp}=0 \label{G8}%
\end{equation}
where $\Omega_{e}=\left(  4\pi e^{2}n_{R}/m_{e}\gamma_{F}\right)  ^{1/2}$ \ is
the frame independent plasma frequency, and is a Lorentz scalar. In the
standard literature, it is conventional to define the frequency $\ \omega
_{e0}=\left(  4\pi e^{2}N_{0}/m_{e}\right)  ^{1/2}\ \ $corresponding to the
density in the lab. frame; the latter is related to the frame- invariant
frequency through$\ \ \ \ \Omega_{e}=\omega_{e0}$\ $\left(  n/\gamma
_{F}\right)  ^{1/2}$\ i.e., the invariant frequency, in this case, is seen as
the lab frame frequency reduced by the relativistic Fermi effects. There are
lots of subtle considerations in creating a strictly Lorentz invariant theory
and the reader is referred to [\onlinecite{SM03}]-[\onlinecite{SMFA17}] for a deeper discussion.

The system of equations describing the longitudinal motion of electron plasma
is given by%

\begin{equation}
\frac{\partial}{\partial t}\gamma_{F}p_{z}+c\frac{\partial}{\partial z}\left(
\gamma\gamma_{F}-\varphi\right)  =0 \label{G9}%
\end{equation}%
\begin{equation}
\frac{\partial}{\partial t}\gamma n+c\frac{\partial}{\partial z}\left(
np_{z}\right)  =0 \label{G10}%
\end{equation}%
\begin{equation}
\frac{\partial^{2}\varphi}{\partial z^{2}}=\frac{\omega_{e0}^{2}}{c^{2}%
}(n\gamma-1) \label{G11}%
\end{equation}
We next carry out a stability analysis for the circularly polarized EM waves.
The monochromatic pump EM wave with frequency $\omega_{0}$ and wave number
$k_{0}$ , is described by%

\begin{equation}
\mathbf{A}_{\perp}=\frac{1}{2}(\widehat{\mathbf{x}}+i\widehat{\mathbf{y}%
})A\exp(-i\omega_{0}t+ik_{0}z)+c.c. \label{G12}%
\end{equation}
Here $\widehat{\mathbf{x}}$ and $\widehat{\mathbf{y}}$ are the unit vectors,
$A=a\exp(i\psi)$ where $a$ and $\psi$ are real valued amplitude and phase. The
unperturbed state of the plasma is characterized by a purely transverse EM
mode with constant amplitude $a=a_{0}$ ($p_{z}=0=\varphi$) and a constant
comoving density of electrons $n_{0}$. This density is related to the lab
frame density by the relation $n_{0}=1/\gamma_{0}$ (in unites $n_{0}%
=N_{0}/\gamma_{0}$) where $\gamma_{0}\mathbf{=}\sqrt{1+a_{0}^{2}/\gamma
_{F0}^{2}}$ and $\gamma_{F0}=\sqrt{1+\left(  R_{0}/\gamma_{0}\right)  ^{2/3}}$.

The dispersion relation that follows from Eqs.(\ref{G8})-(\ref{G12}) reads%

\begin{equation}
\omega_{0}^{2}=k_{0}^{2}c^{2}+\Omega_{e0}^{2} \label{G13}%
\end{equation}
where $\Omega_{e0}=\omega_{e0}\left(  n_{0}/\gamma_{F0}\right)  ^{1/2}%
=\omega_{e0}/\gamma_{m}^{1/2}$ is the relativistically modified electron
plasma frequency and $\gamma_{m}=\gamma_{0}\gamma_{F0}$ . The expression for
$\gamma_{m}$ can be written as $\gamma_{m}=\sqrt{1+p_{F0}^{2}+a_{0}^{2}}$
where $p_{F0}=\left(  R_{0}/\gamma_{0}\right)  ^{1/3}$ is the Fermi momentum.
In the weakly degenerate case when $R_{0}\ll1$ and $p_{F0}\ll a_{0}$ the Eq.
(\ref{G13}) coincides with the dispersion relation for cold classical plasma
with $\Omega_{e0}=\omega_{e0}/\left(  1+a_{0}^{2}\right)  ^{1/2}$, while for
the arbitrary $R_{0}$ and in absence of pump wave ($a_{0}=0$) \ the modified
plasma frequency reads $\Omega_{e0}=\omega_{e0}/\left(  1+R_{0}^{2/3}\right)
^{1/2}$.

Most spectacular manifestations of plasma degeneracy, however, occur for
radiation with relativistically large amplitudes, i.e., when $a_{0}>>1$, and
the radiative modification of the effective electron- mass becomes comparable
to or dominant over the degeneracy modification (see [\onlinecite{SMFA16}]) for a
detailed discussion on radiative renormalization of the electron mass). For
extreme relativistic amplitudes, the effective $\gamma$ simplifies to
$\gamma_{m}\simeq\sqrt{1+a_{0}^{2}}$, essentially the expression for a non
degenerate plasma. For arbitrary values of $a_{0}$ and $R_{0}$ (measuring the
degeneracy strength), the relativistic factor $\gamma_{0}$ can be found from
the following implicit relation:

\begin{equation}
\left(  \gamma_{0}^{2}-1\right)  \left(  1+\left(  \frac{R_{0}}{\gamma_{0}%
}\right)  ^{2/3}\right)  =a_{0}^{2} \label{G14}%
\end{equation}
that can be readily solved. In Fig.1, we plot $\gamma_{0}$ vs $a_{0}$ for
different values of $R_{0}$. Dashed part of each curve corresponds to
$p_{F0}<1$ . For larger values of $R_{0}$, stronger fields are required to
achieve large values of $\gamma_{0}$.

\begin{figure}[h!]
\centering
\includegraphics[scale=1.0]{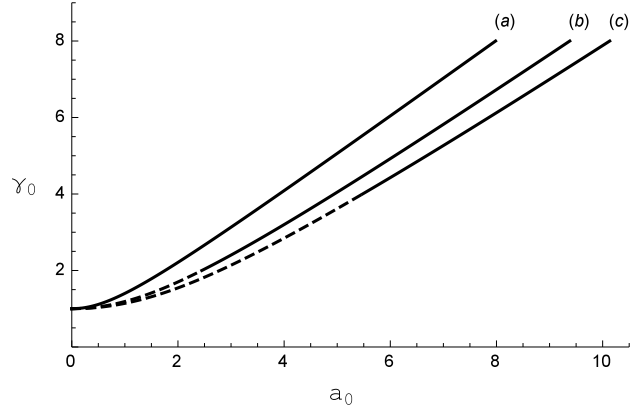}
\caption{\label{pl1capt}Dependence of $\gamma_{0}$ on the EM field strength $a_{0}$ for different level of degeneracy: (a) $R_{0}=0.01$, (b) $R_{0}=2$, and (c) $R_{0}=4$.}
\end{figure}

A short message of the preceding results is that in the field of strong EM
waves (see Fig.1), the effects of electron degeneracy become subdominant and
the plasma responds as a weakly degenerate system.

To investigate the stability of ground wave solution we introduce the small
perturbations in the system $f=f_{0}+\delta f$ ($f_{0}\gg\delta f$) where
$f=\left(  a,\psi,n,p_{z,}\varphi\right)  $. Neglecting higher order terms,
Eqs.(\ref{G8})-(\ref{G11}) can be reduced to the following system of coupled equations:%

\[
\left(  2\omega_{0}\frac{\partial}{\partial t}+2k_{0}c^{2}\frac{\partial
}{\partial z}\right)  \delta\psi+\left(  \frac{\partial^{2}}{\partial t^{2}%
}-c^{2}\frac{\partial^{2}}{\partial z^{2}}\right)  \frac{\delta a}{a_{0}}+
\]

\begin{equation}
+\Omega_{e0}^{2}\left(  \frac{\delta n}{n_{0}}-\frac{\delta\gamma_{F}}%
{\gamma_{F0}}\right)  =0 \label{G15}%
\end{equation}

\begin{equation}
\left(  2\omega_{0}\frac{\partial}{\partial t}+2k_{0}c^{2}\frac{\partial
}{\partial z}\right)  \frac{\delta a}{a_{0}}+\left(  c^{2}\frac{\partial^{2}%
}{\partial z^{2}}-\frac{\partial^{2}}{\partial t^{2}}\right)  \delta\psi=0
\label{G16}%
\end{equation}%
\begin{equation}
\left[  \frac{\partial^{2}}{\partial t^{2}}+\Omega_{e0}^{2}\right]  \left(
\frac{\delta\gamma}{\gamma_{0}}+\frac{\delta n}{n_{0}}\right)  =c^{2}%
\frac{\partial^{2}}{\partial z^{2}}\left(  \frac{\delta\gamma}{\gamma_{0}%
}+\frac{\delta\gamma_{F}}{\gamma_{F_{0}}}\right)  \label{G17}%
\end{equation}

while relations between $\delta\gamma$ and $\delta\gamma_{F}$ obtained from
Eqs.(\ref{G6})-(\ref{G7}) are

\begin{align}
\frac{\delta\gamma}{\gamma_{0}}  &  =\alpha\left(  \frac{\delta a}{a_{0}%
}-\frac{\delta\gamma_{F}}{\gamma_{F_{0}}}\right) \nonumber\\
\frac{\delta\gamma_{F}}{\gamma_{F_{0}}}  &  =\beta\frac{\delta n}{n_{0}}
\label{G18}%
\end{align}
with $\alpha=(1-1/\gamma_{0}^{2})$ and $\beta=(1-1/\gamma_{F_{0}}^{2})/3$.

To obtain the dispersion equation, we assume that all perturbations depend on
the coordinates and time like $\delta f\sim\exp(-i\omega t+ikz)$. After some
algebra the Eqs.(\ref{G15})-(\ref{G18}) leads to dispersion relation%

\begin{equation}
\frac{\Omega_{e0}^{2}}{2}A_{a}\frac{1}{D_{e}}\left(  c^{2}k^{2}+\Omega
_{e0}^{2}-\omega^{2}\right)  \left(  \frac{1}{D_{-}}+\frac{1}{D_{+}}\right)
=1 \label{G19}%
\end{equation}
where $D_{\pm}=(\omega_{0}\pm\omega)^{2}-c^{2}(k_{0}\pm k)^{2}-\Omega_{e0}%
^{2}$ and $D_{e}=\omega^{2}-\Omega_{e0}^{2}-V_{m}^{2}k^{2}c^{2}$. Here
$V_{m}^{2}=\beta\left(  1-\alpha\right)  /\left(  1-\alpha\beta\right)  $,
$\ A_{a}=\alpha\left(  1-\beta\right)  /\left(  1-\alpha\beta\right)  $ which
can be presented in the explicit form as%

\begin{equation}
V_{m}^{2}=\frac{1}{3}V_{F0}^{2}\left(  1-\frac{a_{0}^{2}}{\gamma_{m}^{2}%
}\right)  \left(  1-\frac{1}{3}V_{F0}^{2}\frac{a_{0}^{2}}{\gamma_{m}^{2}%
}\right)  ^{-1} \label{G20}%
\end{equation}

\begin{equation}
A_{a}=\frac{a_{0}^{2}}{\gamma_{m}^{2}\gamma_{m0}^{2}}\label{G21}%
\end{equation}
where $V_{F0}$ is the dimensionless ($V_{F0}\rightarrow V_{F0}/c$) \ Fermi
velocity defined by the relation $V_{F0}=\left(  p_{F0}/\gamma_{F0}\right)  $
and $\gamma_{m0}=1/\sqrt{1-V_{m}^{2}}$.

The dispersion relation (\ref{G19}) has the form similar to the one obtained
in [\onlinecite{Mima}]. In this study the authors investigated the SRS in classical
relativistic plasmas where thermal effects were included by a waterbag model
for electron distribution. It so happens that the description of a fully
degenerate plasma is formally analogous to a waterbag model [\onlinecite{Korakis}]
 and
this analogy is reflected in the form of Eq.(\ref{G19}). Important
differences, induced by plasma degeneracy, are contained in the expressions
(\ref{G20})-(\ref{G21}) \. Note that in classical plasma $V_{m}$ \ is
determined by the temperature (an independent parameter of the problem) while
in a degenerate system, $V_{m}$ depends on plasma density ( determining the
Fermi "temperature"). For ultra-relativistic degeneracy ($R_{0}>>1$),
\ $V_{m}\rightarrow1/\sqrt{3}$. Hence, the maximal value of the parameter
$\gamma_{m0}\simeq1.22.$

The solutions of equation (\ref{G19}) lead to resonance backward and forward
SRS. Like in classical plasma, under certain simplified assumptions, temporal
growth rates ($\Gamma=\operatorname{Im}(\omega)$) of instabilities can be
obtained analytically. However, since the dispersion relation is a sixth order
algebraic equation for $\omega$, numerical solutions may be more useful.

The transverse EM, and the plasma waves are coupled by the parameter
$\chi=\Omega_{e0}^{2}A_{a}$ and the growth rates of above mentioned
instabilities are proportional to a certain power of the coupling parameter
$\chi$. In weakly degenerate case $R_{0}\ll1$ and finite strength of the field
amplitude ($a_{0}$) $\gamma_{m}\approx\sqrt{1+a_{0}^{2}},V_{m}\ll1$ and
$\chi=\omega_{e0}^{2}/\gamma_{m}^{3}$. In this case the dispersion relation
(\ref{G19}) and, consequently, the results of instability coincide with the
results obtained in classical cold plasma embedded in the field of arbitrary
strong radiation [\onlinecite{Mori}],[\onlinecite{Kirsanov}]. Here we would like to remark
that our consideration is valid if the average kinetic energy of electrons
($\sim\epsilon_{F}$) is larger than their interaction energy ($\sim e^{2}%
N_{0}^{1/3}$). This condition is fulfilled for a sufficiently dense plasma
when $N_{0}\gg10^{23}cm^{-3}$ ($1\gg R_{0}\gg10^{-7}$).

The main features and peculiarities of SRS instability in plasma with finite
level of degeneracy parameter $R_{0\text{ }}$can be deduced by analyzing Eqs.
(\ref{G20})-(\ref{G21}) for different limiting cases. We look for the solution
of Eq.(\ref{G19}) in the form $\omega=\omega_{L}+\delta\omega$, where
$\omega_{L}=\sqrt{\Omega_{e0}^{2}+V_{m}^{2}k^{2}c^{2}}$. \ The maximum growth
rates are obtained when scattered wave is also resonant $D_{-}\left(
\omega=\omega_{L}\right)  =0$ leading to the following relation:%
\begin{equation}
\omega_{0}=\left(  \Omega_{e0}^{2}+c^{2}k^{2}V_{m}^{2}\right)  ^{1/2}+\left(
\left(  k_{0}-k\right)  ^{2}+\Omega_{e0}^{2}\right)  ^{1/2} \label{G22}%
\end{equation}

Using Eq.(\ref{G22}) one can show that in an underdense plasma ($\omega
_{0}>2\Omega_{e0}$) the wave vector of resonant modes $k$ lies in the range
$0<k<2k_{0}$. The modes with $k>k_{0}$ and $k<k_{0}$ lead respectively to
backward and forward Raman instabilities. For highly transparent plasma
($\omega_{0}\gg\Omega_{e0}$) the backward Raman instability develops at
$k\simeq2\omega_{0}/c\left(  1+V_{m}\right)  $ and $\omega_{L}=\left[
\Omega_{e0}^{2}+4\omega_{0}^{2}V_{m}^{2}/\left(  1+V_{m}\right)  ^{2}\right]
^{1/2}$ [\onlinecite{Mima}]. Neglecting the nonresonant term ($\sim1/D_{+}$) in
Eq.(\ref{G19}) and making approximations $D_{-}=-2\left(  \omega_{0}%
-\omega_{L}\right)  \delta\omega$ and $\omega_{L}>>\delta\omega$ (and
recalling that $\Omega_{e0}=\omega_{e0}/\gamma_{m}^{1/2}$) for the growth rate
we get the following expression:%

\begin{equation}
\Gamma=\frac{\omega_{e0}}{\sqrt{2}}\frac{a_{0}}{\gamma_{m}^{3/2}}\frac
{\omega_{0}\left(  1-V_{m}\right)  }{\sqrt{\left(  \omega_{0}-\omega
_{L}\right)  \omega_{L}}} \label{G23}%
\end{equation}
For forward Raman scattering instability $k\ll k_{0}\approx\omega_{0}/c$ both
downshifted and upshifted scattered waves are resonant modes $D_{\pm}%
=2(\omega_{L}\pm\omega_{0})\delta\omega$. The maximum growth rate,%

\begin{equation}
\Gamma=\frac{\omega_{e0}^{2}}{2\omega_{0}}\frac{a_{0}}{\gamma_{m}^{2}%
\gamma_{m0}} \label{G24}%
\end{equation}
occurs at $k=\Omega_{e0}\gamma_{m0}/c$ and $\omega_{L}=\Omega_{e0}\gamma_{m0}$.

As per expectations, in the limit of nonrelativistic degeneracy ($R_{0}\ll1$),
Eqs.(\ref{G23})-(\ref{G24}) tend to the classical results for a cold plasma.
For finite level of degeneracy parameter $R_{0}$, and nonrelativistic
strengths of the field ($a_{0}\ll1$), the growth rates of forward Raman
instability is reduced by factor$\left(  1+R_{0}^{2/3}\right)  $in comparison
to weakly degenerate case, while for backward instability the reduction factor
turns out to be $\left(  1+R_{0}^{2/3}\right)  ^{3/4}$.

Most interesting regime, explored in this note, is that of a relativistic
degenerate plasma $\left(  R_{0}\geq1\right)  $embedded in the field of a
relativistically strong EM wave ($a_{0}\gg1$). In the regime of very strong
radiation ($a_{0}/\gamma_{0}\gg1$(see Fig.1 and comments after Eq.(\ref{G13}%
)), even a highly degenerate plasma responds as a weakly degenerate one. For
ultrarelativistic amplitudes, defined by $\ R_{0}/\gamma_{0}\ <1\ $, the Fermi
velocity becomes small $\left(  V_{m}\rightarrow0\right)  ,$and $\gamma
_{m}\approx\sqrt{1+a_{0}^{2}}$. Consequently, the character of the SRS
instability, in particular its growth rate, tends to approach the cold
classical plasma results.

In this letter, we investigated the linear stage of SRS\ instability of a
propagating arbitrary amplitude circularly polarized EM wave in a degenerate
electron plasma. For 1-D wave propagation, it is demonstrated that in the
field of ultra-relativistic amplitude waves, the relativistic degenerate
plasma effectively responds as a weakly degenerate plasma due to reduction of
Fermi momentum. While for weak, nonrelativistic amplitudes (but relativistic
degenerate plasma) the growth rates of instability reduce with increasing density.

The elucidation of the SRS instability induced by ultra strong EM waves in a
relativistic degenerate plasma is highly pertinent to understanding the
dynamics of $X$-ray pulses emanating from compact astrophysical objects. It
will be equally relevant for an exploration of nonlinear interactions between
intense laser pulses and a dense degenerate plasma; the latter class of
physical systems are likely to be realized in the next-generation intense
laser solid density plasma experiments.

The research was supported by the Shota Rustaveli National Science Foundation
grant (DI-2016-14). The research of GT was supported by the Knowledge
Foundation at the Free University of Tbilisi.

\end{document}